\documentclass[aps,prb,twocolumn,superscriptaddress,showpacs]{revtex4-1}

\usepackage{graphicx}
\usepackage{color}

\newcommand{\CCO}{CuCrO$_{2}$}
\newcommand{\CCMO}{CuCr$_{0.97}$Mg$_{0.03}$O$_{2}$}
\newcommand{\CACO}{Cu$_{0.85}$Ag$_{0.15}$CrO$_{2}$}
\newcommand{\CCAO}{CuCr$_{0.85}$Al$_{0.15}$O$_{2}$}

\begin{document}

\title{Influence of Mg, Ag and Al substitutions on the magnetic
excitations \\
in the triangular-lattice antiferromagnet CuCrO$_{2}$}

\author{R. Kajimoto}
\email[E-mail: ]{ryoichi.kajimoto@j-parc.jp}
\altaffiliation[On loan from ]{J-PARC Center, Japan Atomic Energy Agency,
Tokai, Ibaraki 319-1195, Japan}
\affiliation{Research Center for Neutron Science and Technology,
Comprehensive Research Organization for Science and Society (CROSS),
Tokai, Ibaraki 319-1106, Japan}

\author{K. Nakajima}
\affiliation{J-PARC Center, Japan Atomic Energy Agency, Tokai, Ibaraki
319-1195, Japan}

\author{S. Ohira-Kawamura}
\affiliation{J-PARC Center, Japan Atomic Energy Agency, Tokai, Ibaraki
319-1195, Japan}

\author{Y. Inamura}
\affiliation{J-PARC Center, Japan Atomic Energy Agency, Tokai, Ibaraki
319-1195, Japan}

\author{K. Kakurai}
\affiliation{Quantum Beam Science Directorate, Japan Atomic Energy
Agency, Tokai, Ibaraki 319-1195, Japan}


\author{T. Hokazono}
\affiliation{Department of Nano-Structures and Advanced Materials,
Kagoshima University, \\ Kagoshima 890-0065, Japan}

\author{S. Oozono}
\affiliation{Department of Nano-Structures and Advanced Materials,
Kagoshima University, \\ Kagoshima 890-0065, Japan}

\author{T. Okuda}
\affiliation{Department of Nano-Structures and Advanced Materials,
Kagoshima University, \\ Kagoshima 890-0065, Japan}

\date{\today}

\begin{abstract}
Magnetic excitations in CuCrO$_{2}$, CuCr$_{0.97}$Mg$_{0.03}$O$_{2}$,
Cu$_{0.85}$Ag$_{0.15}$CrO$_{2}$, and CuCr$_{0.85}$Al$_{0.15}$O$_{2}$
have been studied by powder inelastic neutron scattering to elucidate
the element substitution effects on the spin dynamics in the Heisenberg
triangular-lattice antiferromagnet CuCrO$_{2}$. The magnetic excitations
in CuCr$_{0.97}$Mg$_{0.03}$O$_{2}$ consist of a dispersive component and
a flat component. Though this feature is apparently similar to
CuCrO$_{2}$, the energy structure of the excitation spectrum shows some
difference from that in CuCrO$_{2}$. On the other hand, in
Cu$_{0.85}$Ag$_{0.15}$CrO$_{2}$ and CuCr$_{0.85}$Al$_{0.15}$O$_{2}$ the
flat components are much reduced, the low-energy parts of the excitation
spectra become intense, and additional low-energy diffusive spin
fluctuations are induced. We argued the origins of these changes in the
magnetic excitations are ascribed to effects of the doped holes or
change of the dimensionality in the magnetic correlations.
\end{abstract}


\maketitle

\section{Introduction}

Two-dimensional (2D) triangular-lattice antiferromagnet is one of the
simplest playgrounds of the geometrical frustration. If Heisenberg spins
on the triangular-lattice layer are coupled by the nearest neighbor
exchange interactions, the frustration is dissolved by forming a
120$^{\circ}$ spin ordering. However, the ground state becomes
nontrivial when there exist finite inter-layer or longer-distance
couplings. In other words, if we can control these extra parameters, we
can possibly create some novel states. Such attempts have recently been
conducted on one of the model materials of the triangular-lattice
Heisenberg antiferromagnets CuCrO$_{2}$. In this compound,
triangular-lattice layers of magnetic Cr$^{3+}$ ions (3d$^{3}$) with $S
= 3/2$ are separated from each other by non-magnetic layers of Cu$^{+}$
ions (3d$^{10}$). With decreasing temperature ($T$), the Cr spins form
the three-dimensional (3D) magnetic ordering due to finite inter-layer
couplings. The magnetic structure is a proper screw structure whose wave
vector is $(q, q, 0)$ with $q$ being slightly smaller than the value of
the 120$^{\circ}$ structure, $q =
1/3$.\cite{kadowaki90,poienar90,soda09} Recently, it was found that
element substitution can induce interesting magnetic and transport
properties. For example, substitution of the intra-layer Cr ions with
non-magnetic Mg ions (CuCr$_{1-x}$Mg$_{x}$O$_2$) slightly increases the
transition temperature of the 3D magnetic ordering ($T_N$), and sharpens
the magnetic transition. Since the promotion of the 3D magnetic ordering
is followed by increase of the magnetization, appearance of the
magnetoresistance, and drastic enhancement of the electric conductivity,
it was interpreted as the effect of holes introduced by the substitution
of Cr$^{3+}$ ions by Mg$^{2+}$ ions.\cite{okuda05,okuda08} On the other
hand, substitution of the inter-layer Cu$^{+}$ ions with Ag$^{+}$ ions
(4d$^{10}$) (Cu$_{1-y}$Ag$_{y}$CrO$_{2}$) and that of intra-layer
Cr$^{3+}$ ions with non-magnetic Al$^{3+}$ ions
(CuCr$_{1-x}$Al$_{x}$O$_{2}$) make the 3D magnetic ordering unclear. In
both the systems, $T^3$ dependence of the low-temperature magnetic
specific heat ($C_\mathrm{mag}$) of {\CCO} is replaced by the $T^2$
dependence as the substitution is progressed, indicating the development
of the two-dimensionality in the magnetic correlations. Moreover, close
investigations of the $C_\mathrm{mag}$'s and corresponding entropies
showed that development of some unconventional low-energy spin
fluctuations.\cite{okuda09,okuda11}

With these findings, we can expect that the element substitutions also
affect the spin dynamics. Elucidation of this change in the spin
dynamics should be important information to understand the origin of the
physical properties of the substituted compounds and an important clue
to produce novel phenomena in geometrically frustrated magnets by the
element substitution. Accordingly, we performed inelastic neutron
scattering study on powder samples of Mg, Ag, and Al substituted {\CCO}
in addition to {\CCO} to observe their magnetic excitations as functions
of momentum and energy transfers. We found that a variety of the
magnetic excitations are induced by different types of the element
substitutions.

\section{Experiments}

About 3~cm$^3$ polycrystalline samples of {\CCO}, {\CCMO}, {\CACO}, and
{\CCAO} were synthesized by a standard solid-state reaction
method.\cite{okuda09,okuda11,okuda05,okuda08} They have rhombohedral
crystal structures (space group $R\bar{3}m$) with the lattice constants
$a \sim 3.0$~{\AA} and $c \sim 17$~{\AA}. Their $T_N$'s determined by
specific heat measurements are 24~K (CuCrO$_2$), 13~K [{\CACO}], 26~K
[{\CCMO}] and 24~K [{\CCAO}].\cite{okuda11} The neutron scattering
measurements were performed with the time-of-flight technique using the
cold neutron disk-chopper spectrometer AMATERAS at the Materials and
Life Science Experimental Facility (MLF) in the Japan Proton Accelerator
Research Complex (J-PARC).\cite{amateras} The power of the accelerator
was $\sim$120~kW. We utilized multiple incident energies of neutrons,
$E_i$ = 15, 7.7, 4.7, and 3.1~meV, which were produced simultaneously
taking advantage of the repetition rate multiplication by the
monochromating disk chopper.\cite{nakamura07} We analyzed all the data
with the software suite \textsc{utsusemi}\cite{utsusemi} to obtain the
powder averaged scattering function $S(Q,E)$ ($Q$ is the amplitude of
momentum transfer and $E$ is the energy transfer) as described in
Ref.~\onlinecite{kajimoto10}. The difference in utilized neutron flux
between data was normalized versus the proton beam current incident on
the neutron target. The difference in volume between the samples was
normalized by intensities of nuclear Bragg peaks [{\CCO} and {\CACO}] or
the mole numbers [{\CCO}, {\CCMO} and {\CCAO}]. The preliminary results
of the present study have been reported in
Ref.~\onlinecite{kajimoto_LT}.

\section{Results}

\subsection{{\CCO} and {\CCMO}}

\begin{figure}[t]
 \includegraphics[width=0.9\hsize]{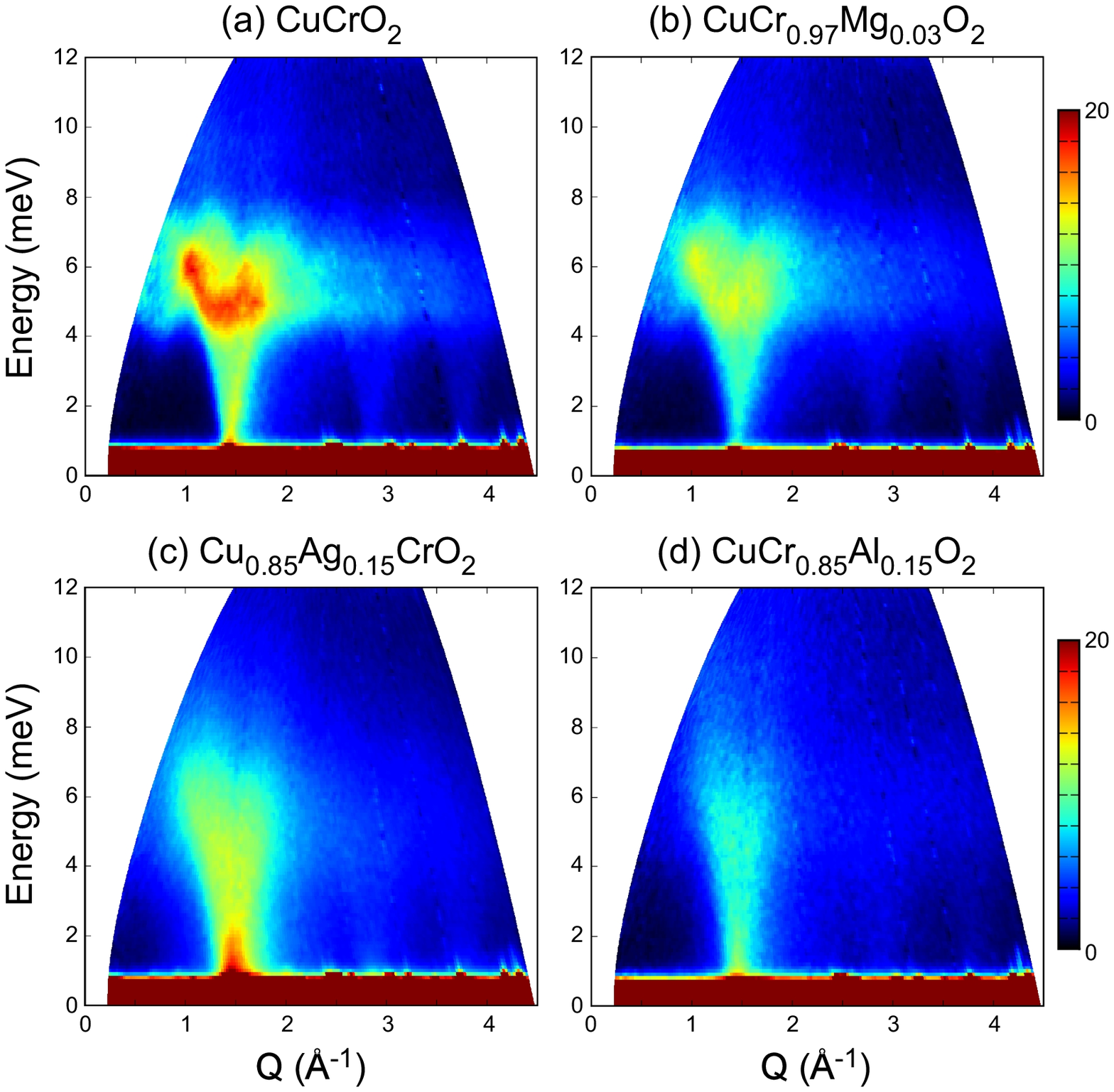}
 \caption{(Color online) $Q$-$E$ maps of the excitation spectra of (a)
 {\CCO}, (b) {\CCMO}, (c) {\CACO}, and (d) {\CCAO} measured at 6, 5, 5
 and 4~K, respectively, with $E_i = 15$~meV.\label{QEmapsEi15meV}}
\end{figure}

\begin{figure}[t]
 \includegraphics[scale=0.5]{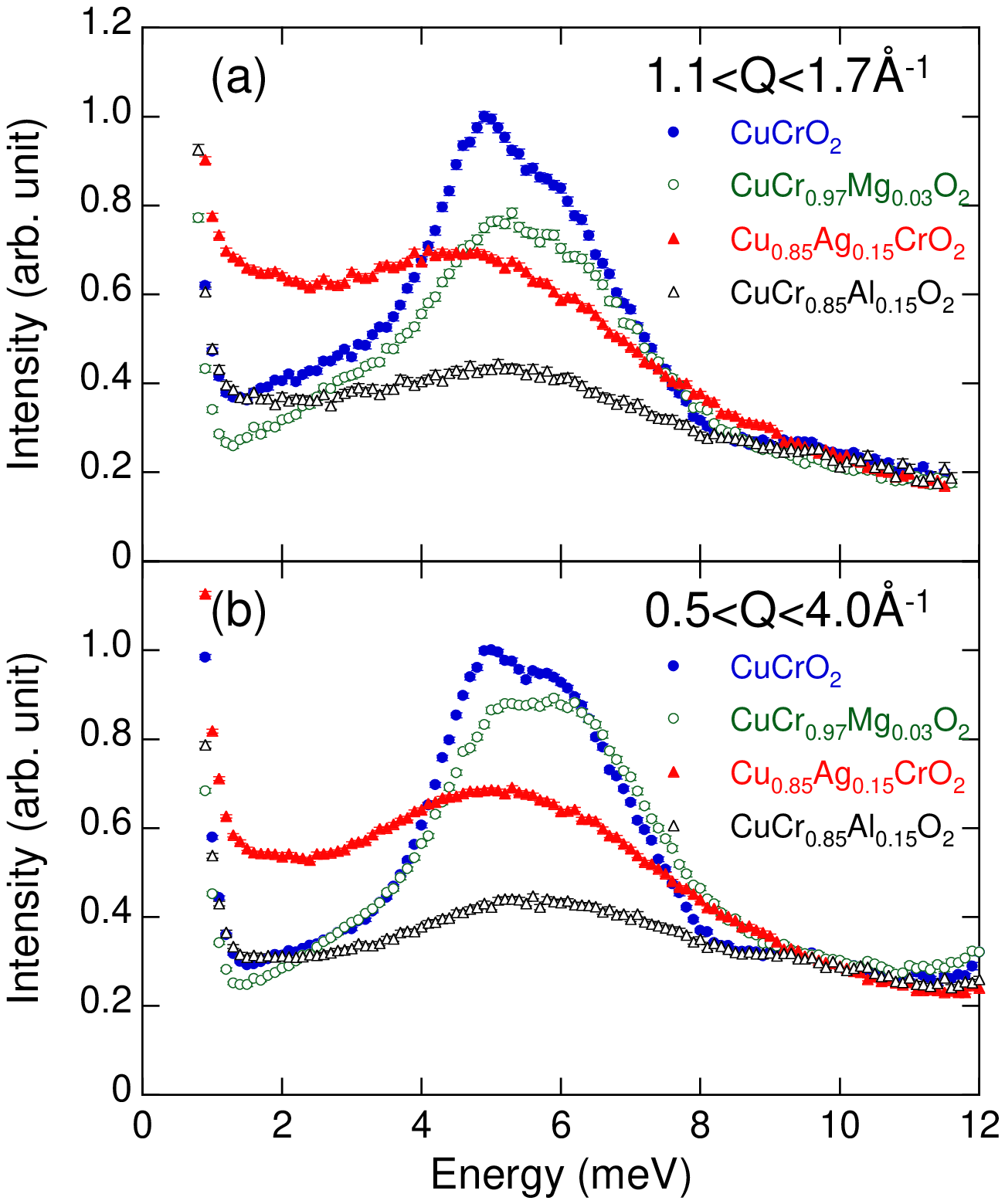}
 \caption{(Color online) Energy dependence of the excitation spectra of
 {\CCO} (closed circles), {\CCMO} (open circles), {\CACO} (closed
 triangles), and {\CCAO} (open triangles) measured at 6, 5, 5 and 4~K,
 respectively, with $E_i = 15$~meV. The data are obtained by integrating
 the data shown in Fig.~\ref{QEmapsEi15meV} over $Q$ ranges of (a)
 1.1--1.7~\AA$^{-1}$ and (b) 0.5--4.0~\AA$^{-1}$. The vertical axes are
 scaled so that peaks of the {\CCO} data at $E \sim 5$~meV become
 unity. \label{EprofEi15meV}}
\end{figure}

First, we show results for the two compounds which show clear 3D
ordering behavior, {\CCO} and {\CCMO}. Figure~\ref{QEmapsEi15meV}(a)
shows the $S(Q,E)$ map of {\CCO} measured at $T \sim 5$~K. It shows well
defined magnetic excitations in particular around $Q = 1.4$~\AA$^{-1}$,
whose value corresponds to the wave vector of the magnetic ordering
$\mathbf{Q}_m \sim (1/3,1/3,0)$. The magnetic excitations in {\CCO}
consist of two kinds of components: one is a steep dispersive component
at $Q_m \sim 1.4$~\AA$^{-1}$. The other is a less dispersive component
(flat component) at $E \sim 5$~meV spreading over a wide range of
$Q$.\cite{kajimoto10} By comparing our data with the magnetic excitation
spectra for single crystal samples,\cite{frontzek11,poienar10} the
dispersive component and the flat component correspond to the $\alpha$
and $\beta$ modes in Ref.~\onlinecite{frontzek11}, respectively.  The
flat component forms a large peak at $E \sim 5$~meV in the energy
profile as shown by closed circles in Fig.~\ref{EprofEi15meV}, and the
dispersive component shows a tail at lower energies whose intensity
decreases with decreasing the energy. In addition, there exist weak but
finite excitations at a higher-energy region than the flat component, as
manifested by diffusive signal around 10~meV in
Fig.~\ref{QEmapsEi15meV}(a) and a tail in the energy profile in
Fig.~\ref{EprofEi15meV}. They may correspond to the $\beta'$ mode in
Ref.~\onlinecite{frontzek11}.

{\CCMO} shows apparently similar excitation spectrum to that of {\CCO}:
It consists of a dispersive component and a substantial flat component
[Fig.~\ref{QEmapsEi15meV}(b)], and its energy profile shows a large peak
at $E \sim 5$~meV (open circles in Fig.~\ref{EprofEi15meV}). However,
the overall weight of the excitation spectrum of {\CCMO} is smaller than
that of {\CCO}. The reduction of the spectral weight may partly be
attributed to disorder introduced by the Mg substitution, which was
confirmed by recent $\mu$SR\cite{ikedo09} and thermal conductivity
studies\cite{okuda_un}. However, we found that the reduction in spectral
weight depends on energy, and it is especially evident around the peak
at $E \sim 5$~meV as well as in a low-energy part of the spectrum at $E
< 4$~meV. Though the reduction at these energies is particularly evident
in the energy profile around $Q_m \sim 1.4$~\AA$^{-1}$
[Fig.~\ref{EprofEi15meV}(a)], it remains even if we integrate the
excitation spectra over as wide a $Q$ region as possible
[Fig.~\ref{EprofEi15meV}(b)]. This means that the reduction of the
spectral weight cannot be explained only by the broadening in $Q$ by the
disorder. In fact, the $Q$ profile of the low-energy part of the
excitation spectra in {\CCMO} has almost similar width to that in {\CCO}
[Fig.~\ref{EprofQprofEi4.7meV}(a)]. On the other hand, {\CCMO} has a higher
weight in a higher-energy region (at $E \sim 8$~meV). These facts show that
the Mg substitution does not only induces reduction of the spectral
weight but also induces a change of the energy structure of the magnetic
excitations.

\begin{figure}
 \includegraphics[scale=0.5]{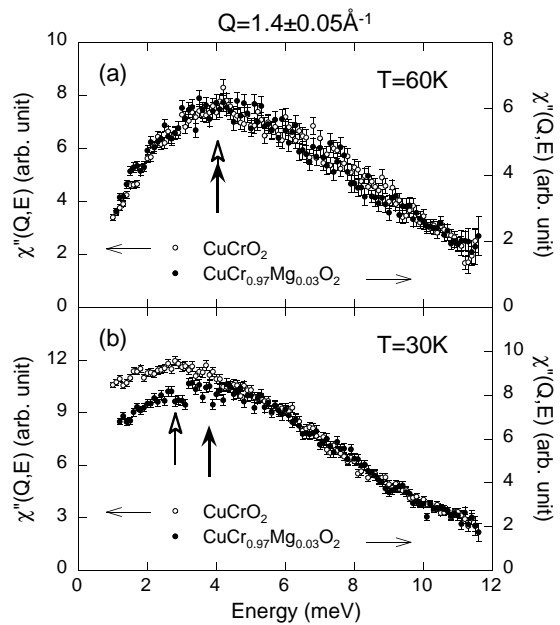}
 \caption{$\chi''(Q,E)$ of {\CCO} (open circles) and
 {\CCMO} (closed circles) at $Q = 1.4\pm
 0.05$~\AA$^{-1}$ measured at (a) 60~K and (b) 30~K, which are
 deduced from data with $E_i = 15$~meV. White and black arrows
 indicate the peak positions in $\chi''(Q,E)$ of {\CCO} and
 {\CCMO}, respectively. \label{CCO_CCMO_30K_60K}}
\end{figure}

The similarity in the slope of the dispersive component between {\CCO}
and {\CCMO} [Figs.~\ref{QEmapsEi15meV}(a) and \ref{QEmapsEi15meV}(b)]
suggests that the exchange interactions between the nearest Cr sites,
$J_1$, is almost the same in the both compounds. On the other hand, the
difference in the weight of the flat component suggests that the there
is some difference in the exchange interactions between the second
nearest neighbor Cr sites, $J_2$, since the energy of the flat mode
($\beta$ mode) is determined by $J_2$.~\cite{frontzek11} This argument
of $J_1$ and $J_2$ is also supported by the magnetic excitations at $T >
T_N$. Since the magnetic correlations start to develop far above $T_N$
for the both compounds,\cite{kadowaki90,okuda05,li11} they show clear
magnetic excitations at $Q \sim 1.4$~\AA$^{-1}$ at $T >
T_N$.\cite{kajimoto10} Figure~\ref{CCO_CCMO_30K_60K} shows the dynamical
susceptibility $\chi''(Q,E) = S(Q,E)/[1-\exp(-E/k_B T)]$ at $Q =
1.4$~\AA$^{-1}$, which are derived from constant-$Q$ cuts of $S(Q,E)$ at
60~K and 30~K. At 60~K, far above $T_N$, the shapes of $\chi''$'s of the
both compounds completely coincides with each other
[Fig.~\ref{CCO_CCMO_30K_60K}(a)]. However, at 30~K, slightly above
$T_N$, the peak position of $\chi''$ of {\CCO} decreases with
approaching the critical divergence, while that of {\CCMO} stays at
almost the same energy as 60~K [Fig.~\ref{CCO_CCMO_30K_60K}(b)]. This
difference in the temperature dependence of $\chi''$ is consistent with
a recent electron spin resonance (ESR) study.\cite{li11} They reported
that the ESR intensity of {\CCO} drops rapidly below 50~K, while that of
CuCr$_{0.98}$Mg$_{0.02}$O$_{2}$ shows a monotonous increase with cooling
and reduces rapidly only below $T_N$. The small temperature dependence
of $\chi''$ of {\CCMO} observed in the present study should be related
to the monotonous increase of the ESR intensity at $T > T_N$ in
Ref.~\onlinecite{li11}. Considering that the peak energy in $\chi''$
(arrows in Fig.~\ref{CCO_CCMO_30K_60K}) is a measure of the energy scale
of spin fluctuations, the difference in the temperature dependence of
$\chi''$ may be attributed to some difference in the magnetic
interactions between the two compounds. Since the contribution of the
largest magnetic exchange interaction $J_1$ to the magnetic excitation
spectrum becomes dominant at high temperatures, the almost identical
$\chi''$'s for the two compounds at 60~K suggest that $J_1$'s in {\CCO}
and {\CCMO} have almost the same values. On the other hand, the
contribution of longer-distance interactions $J_2$ etc. with smaller
magnitudes become effective with decreasing temperature. Therefore, the
difference in $\chi''$ at 30~K suggests the difference in $J_2$ or
longer-distance interactions between the two compounds.

\begin{figure}
 \includegraphics[scale=0.5]{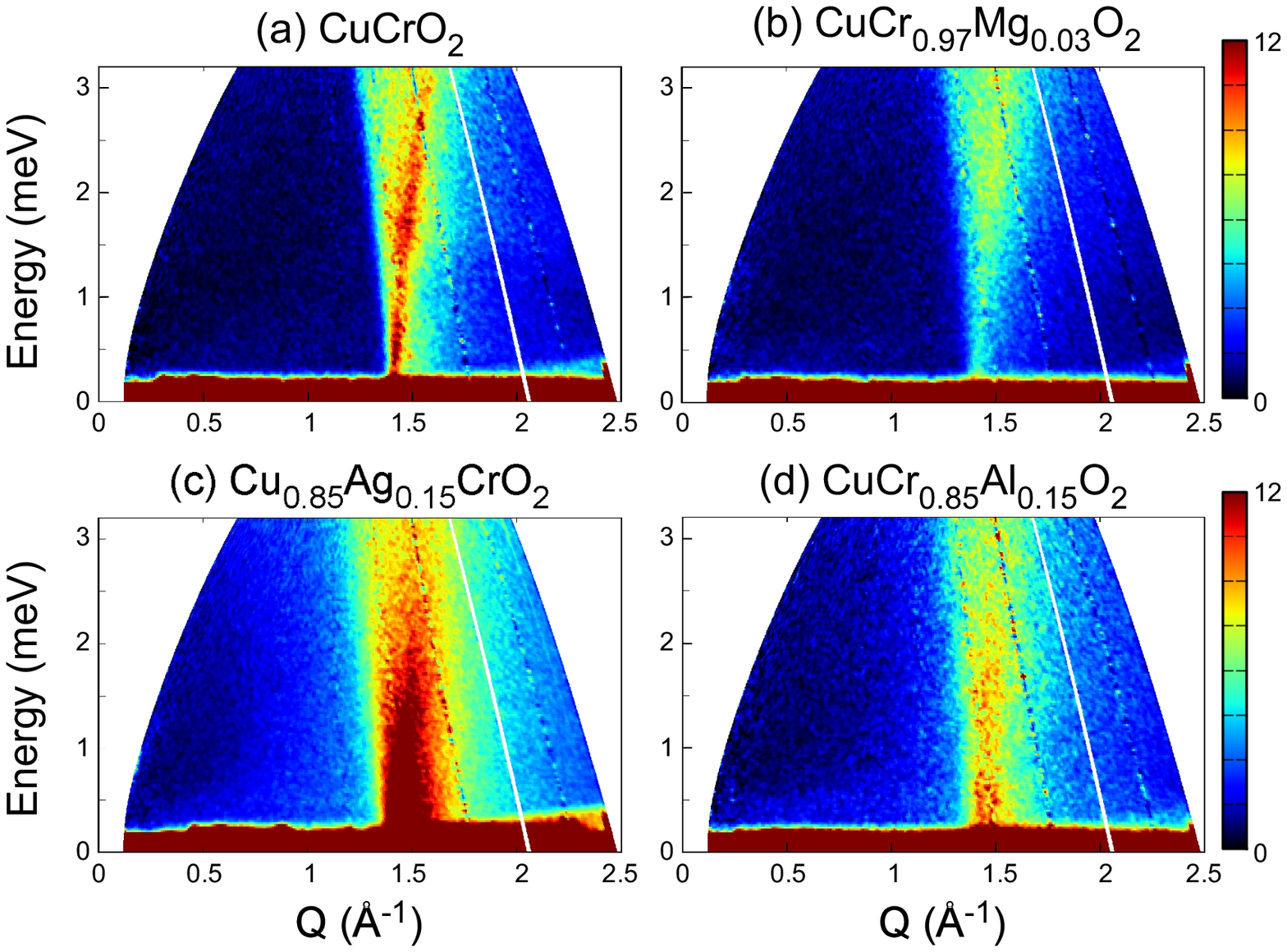}
 \caption{(Color online) $Q$-$E$ maps of the excitation spectra of (a)
 {\CCO}, (b) {\CCMO}, (c) {\CACO}, and (d) {\CCAO} measured at 6, 5, 5
 and 4~K, respectively, with $E_i = 4.7$~meV.\label{QEmapEi4.7meV}}
\end{figure}

\begin{figure}
 \includegraphics[scale=0.5]{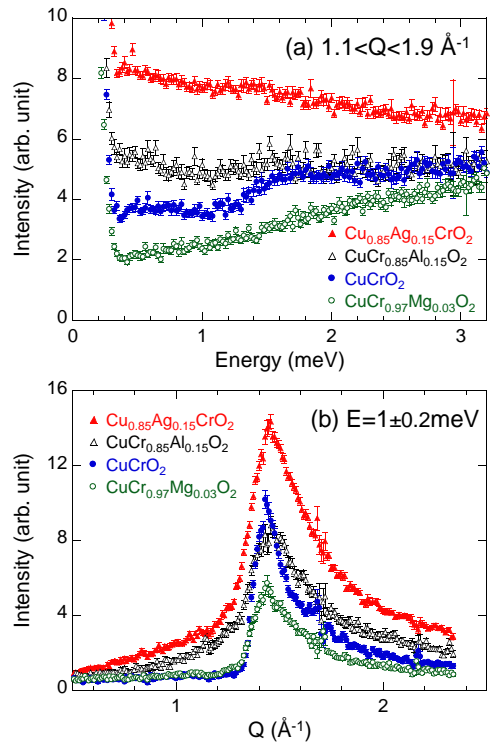}
 \caption{(Color online) (a) Energy dependence of the excitation spectra
 obtained by integrating the data in Fig.~\ref{QEmapEi4.7meV} over $Q$
 ranges of 1.1--1.9~\AA$^{-1}$. (b) Constant-$E$ cuts of the data in
 Fig.~\ref{QEmapEi4.7meV} at $E_i = 1$~meV with a width of $\pm
 0.2$~meV. Closed circles, open circles, closed triangles, and open
 triangles show data of {\CCO}, {\CCMO}, {\CACO}, and {\CCAO},
 respectively.\label{EprofQprofEi4.7meV}}
\end{figure}

In order to see the difference in low-energy part of the excitations in
more detail, we investigated the data with a lower $E_i = 4.7$~meV at $T
\sim 5$~K. The $S(Q,E)$ map of {\CCO} shows a sharp ridge in a broader
dispersive excitations with a tail in the higher $Q$ side
[Fig.~\ref{QEmapEi4.7meV}(a)]. This feature indicates that there exist
clear spin wave excitations, but the inter-layer coupling is much
smaller than the intra-layer coupling.\cite{warren41} By looking at the
excitation spectra carefully, we found that there is a discontinuity in
the ridge excitations around 1~meV. This feature can be seen more
clearly in the energy profile shown by closed circles in
Fig.~\ref{EprofQprofEi4.7meV}(a), where there are shoulders around $E =
0.5$~meV and 1.5~meV. The former should be attributed to the spin gap
which was reported by the previous single crystal
study.\cite{frontzek11} On the other hand, the existence of the shoulder
at $E = 1.5$~meV suggests that there is the second gap in the spin wave
dispersion in {\CCO} around the zone center. Very recently, we confirmed
that the gap at $E = 1.5$~meV in our preliminary single crystal
study. The two gaps indicate there are two kinds of single ion
anisotropies in {\CCO}, which may be ascribed to the easy-axis
anisotropy along the $z$ axis and the easy-plane anisotropy in the
spiral plane.\cite{frontzek11}


On the other hand, in {\CCMO} the spectral weight decreases faster than
that in {\CCO} with decreasing the energy [Figs~\ref{QEmapEi4.7meV}(b)
and \ref{EprofQprofEi4.7meV}(a)]. This energy dependence is quite
interesting considering that generally the structure factor of
antiferromagnetic spin wave is proportional to $1/E$. Another
interesting feature in the spectrum of {\CCMO} in contrast to {\CCO} is
that there is no gap or shoulder structure. These results indicate that
the Mg substitution induces a change in the single-ion anisotropies in
addition to the energy structure of the excitation spectra.

\subsection{{\CACO} and {\CCAO}}

Next, we show results for the rest of the samples, {\CACO} and {\CCAO},
where 3D magnetic transition becomes suppressed and 2D character becomes
evident in the specific heats. In {\CACO} [Fig.~\ref{QEmapsEi15meV}(c)],
the dispersive component becomes diffusive even at temperature far below
$T_N$, and the weight of the excitation spectrum shifts to a lower
energy. Moreover, the flat component decreases its intensity and is hard
to observe. As a result, the low-energy part ($E \lesssim 4$~meV) of the
excitations has similar spectral weight to those at $E \sim 5$~meV with
a steep increase below $E \lesssim 2$~meV (closed triangles in
Fig.~\ref{EprofEi15meV}). These features should be associated with the
increase of the two dimensionality in the magnetic
excitations.\cite{kajimoto10}

The excitation spectrum of {\CCAO} looks similar to {\CACO} in that it
consists of a diffusive dispersive component and a very weak flat
component [Fig.~\ref{QEmapsEi15meV}(d) and open triangles in
Fig.~\ref{EprofEi15meV}].  This fact suggests that the spin dilution in
the Cr layers introduce a similar effect in the magnetic correlations to
that introduced by decrease in the inter-layer coupling by the Ag
substitution. However, {\CCAO} has much smaller spectral weight than
{\CACO}. The ratio of the spectral weight between {\CCAO} and {\CACO} is
independent on the $Q$ region over which the data are integrated
[Figs.~\ref{EprofEi15meV}(a) and \ref{EprofEi15meV}(b)]. This means the
low spectral weight in {\CCAO} is not caused by broadening of the
spectrum in $Q$. Of course, it is partly explained by the 15\% dilution
of Cr spins by the Al substitution. However, the difference between the
two compounds is much larger than 15\%: the intensity of {\CCAO} is
about 60\% of that of {\CACO} at $E \lesssim 5$~meV. The large reduction
of the spin fluctuations may be related to the spin-glass-like behavior
observed in {\CCAO}.\cite{okuda11} We should also note that the
excitation spectrum of {\CCAO} [Fig.~\ref{QEmapsEi15meV}(d)] is
completely different from that of {\CCMO}, which is another evidence
that the difference in the magnetic excitations between {\CCMO} and
{\CCO} can not be explained only by the spin dilution effect.

Figures~\ref{QEmapEi4.7meV}(c) and \ref{QEmapEi4.7meV}(d) show
low-energy parts of excitation spectra of {\CACO} and {\CCAO} around $Q
= Q_m$ measured with $E_i = 4.7$~meV. Their energy dependences obtained
by integrating the data over $Q$ are shown in
Fig.~\ref{EprofQprofEi4.7meV}(a).  The spectral weight of {\CACO}
monotonically increases as the energy decreases
[Fig.~\ref{QEmapEi4.7meV}(c) and closed triangles in
Fig.~\ref{EprofQprofEi4.7meV}(a)]. The spectral weight of {\CCAO} also
shows increase as the energy decreases, though its intensity is lower
than that of {\CACO} [Fig.~\ref{QEmapEi4.7meV}(d) and open triangles in
Fig.~\ref{EprofQprofEi4.7meV}(a)]. The resemblance in the low-energy
spectra of the two compounds is also found in their $Q$
dependences. Figure~\ref{EprofQprofEi4.7meV}(b) shows $Q$ profiles of
the excitation spectra of the four compounds at $E = 1$~meV. In {\CCO},
the profile shows a sharp rise at $Q < Q_m$ and gradual decrease at $Q >
Q_m$, forming a single skewed peak with a long tail at a higher $Q$
region. {\CCMO} also shows a similar single skewed profile, though its
intensity is about half of that of {\CCO}. In contrast to these compound
which form clear 3D magnetic orderings, $Q$ profiles of {\CACO} and
{\CCAO} consist of two components: one is a broad component with less
intensity and long tails even at $Q < Q_m$, and the other is a sharper
component like {\CCO}. As discussed in Ref.~\onlinecite{kajimoto10}, the
broad components manifest the existence of diffusive low-energy spin
fluctuations in addition to disordered spin waves, and they are
attributed to the origin of the $AT^2$ term in the low-temperature
$C_\mathrm{mag}$ with a large value of $A$.

\section{Discussion}

The present study revealed that the Mg substitution induces very
different features in the magnetic excitation spectrum from the Ag and
Al substitutions. While the Mg substitution preserves the sharp
dispersive component at low energies at $Q \sim Q_m$ and flat component
at $E \sim 5$~meV like {\CCO}, the Ag and Al substitutions reduce the
flat component and make the low-energy part of the excitation spectra
diffusive and intense. These difference must be related to the
difference in the dimensionality of the magnetic ordering, which has
been confirmed by the specific heat measurements.\cite{okuda09,okuda11}

Although the magnetic excitations in {\CCO} and {\CCMO} has the above
mentioned similarities, there are clear differences in the excitation
spectrum between {\CCMO} and {\CCO}. The Mg substitution reduces the
spectral weight in particular at low energies and around the flat
component ($E \sim 5$~meV), while it increases a high-energy part ($E
\sim 8$~meV) of the spectrum. There is also difference in temperature
dependence of the magnetic excitations in the paramagnetic phase. In
addition, the two kinds of spin gaps at $Q = Q_m$ disappear. These facts
indicate that the Mg substitution affects the magnetic exchange
interactions, in particular longer-distance interactions than $J_1$, as
well as the single-ion anisotropies. They may cause the slightly higher
$T_N$ and the sharper magnetic transition in {\CCMO} than in {\CCO}.

Though the 3\% spin dilution may induce some disorder in the magnetic
correlations,\cite{ikedo09,okuda_un} it alone cannot explain these
changes, which is also supported by the fact that the spin dilution by
the Al substitution induces completely different effects in the magnetic
excitations. One of the plausible origins of the above changes in the
magnetic excitations is a change in the crystal structure induced by the
3\% substitution of Mg, where the lattice constants show 0.2--0.3\%
increase.\cite{okuda05} However, the increase of the lattice constants
should decrease the antiferromagnetic exchange
interactions,\cite{okuda_takeshita11} which is inconsistent with the
slight increase of $T_N$ by the Mg substitution,\cite{okuda08} and
suggests that the the change in the crystal structure has little effect
on the magnetic exchange interactions. On the other hand, the Mg
substitution increases the magnetization and induces negative
magnetoresistance effect, which were attributed to a coupling between
doped holes and the Cr spins.\cite{okuda05} Then, a more interesting
scenario is that the holes introduced by the Mg substitution and
concomitant increase in the electron conductivity modify the magnetic
excitations. One of the plausible scenarios of the coupling between the
holes and the Cr spins is that ferromagnetic exchange interactions due
to the double-exchange interactions.  They can modify the magnetic
excitations over long distances through the hopping of holes, and this
scenario is consistent with the increase of the magnetization. However,
it is still inconsistent with the increase of $T_N$, because the
ferromagnetic fluctuations should compete with the antiferromagnetic
exchange interactions and lower $T_N$. Okuda \textit{et al.} proposed
the stabilization of the magnetic ordering by the order-by-disorder
mechanism to reconcile the ferromagnetic fluctuations and the increase
of $T_N$.\cite{okuda08,okuda_un} Another interesting scenario related to
the doped holes is the orbital degree of freedom. Since CrO$_{6}$
octahedra of {\CCO} are compressed along the $c$ axis, $t_{2g}$ orbitals
of a Cr ion split into a doubly degenerate states with a higher energy
and a non-degenerate state with a lower energy by the crystal electric
field. In {\CCO}, there are three electrons in $t_{2g}$ orbitals of
Cr$^{3+}$ ions, which results in no orbital degree of freedom. However,
in {\CCMO}, doped holes which enter the doubly degenerate orbitals can
produce the orbital degree of freedom. Therefore, it is plausible that
hopping of the holes induces change of the single-ion anisotropies as
well as exchange interactions between spins through the change of the
orbital states. Though whether this scenario can mainly affect
longer-distance exchange interactions instead of $J_1$ is not clear, it
might be possible depending on the orbital state.

On the other hand, the magnetic excitations of {\CCAO} are very similar
to those of {\CACO} except that their intensity is smaller than those of
{\CACO} by about 60\%. This fact suggests that the both types of the
element substitutions induce similar effects on the magnetic
excitations. In particular, the appearance of additional diffusive
low-energy fluctuations is interesting with expectations for
unconventional spin fluctuations such as spin
liquid\cite{morita02,mizusaki06} or $Z_2$
vortex.\cite{okubo10,kawamura11} In our previous neutron scattering
study on {\CACO}, we interpreted this low-energy spin fluctuation as the
origin of the large $T^2$ component of
$C_\mathrm{mag}$.\cite{kajimoto10} In this study, {\CCAO} also shows
this diffusive low-energy fluctuation, which is consistent to the fact
that CuCr$_{1-x}$Al$_{x}$O$_{2}$ also has the $T^2$ components in its
$C_\mathrm{mag}$.\cite{okuda11} In {\CACO}, the change in the magnetic
excitations should be attributed to the enhancement of the
two-dimensionality by the substitution of the inter-layer Cu ions by Ag
ions. In contrast, the direct effect of the Al substitution is to
disturb the magnetic correlations in the Cr layers. However, since the
3D ordering of a quasi-2D system depends on the development of the
intra-layer correlations in a mean-field approximation,\cite{Jongh_book}
the Al substitution can indirectly weaken the magnetic correlations
between the layers, thereby inducing similar effect on the magnetic
excitations as the Ag substitution.

\section{Conclusion}

We studied magnetic excitations in {\CCO}, {\CCMO}, {\CACO}, and {\CCAO}
by inelastic neutron scattering to elucidate the element substitution
effects on the spin dynamics in the triangular-lattice antiferromagnet
{\CCO}. We found that the magnetic excitations in {\CCO} and {\CCMO}
consist of sharp dispersive components at low energies at $Q \sim Q_m$
and the flat components at $E \sim 5$~meV, while in {\CACO} and {\CCAO}
the former components are diffusive and intense, and the latter
components are much reduced. These difference must be related to the
difference in the character of the dimensionality of the magnetic
correlations. Furthermore, the Mg substitution changes the energy
structure of the excitation spectrum, which suggests some modifications
in the magnetic exchange interactions and single-ion anisotropies by the
doped holes. The Ag and Al substitutions induce additional low-energy
diffusive spin fluctuations, which are likely unconventional spin
dynamics. The present results should be good examples how we can control
the spin dynamics in triangular-lattice antiferromagnets by the element
substitution, and give a hint to find a novel phenomenon in frustrated
low-dimensional systems.

\begin{acknowledgments}
 The experiment on AMATERAS was carried out under the project no.\
 2009A0073. This work was partly supported by Grants-in-Aid for
 Scientific Research on Priority Areas ``Novel States of Matter Induced
 by Frustration'' of the Ministry of Education, Culture, Sports,
 Science, and Technology, Japan (19052001 and 19052004) and by a
 Grant-in-Aid for Scientific Research (B) of the Japan Society for the
 Promotion of Science (20340097).
\end{acknowledgments}

\end{document}